\documentclass[]{aa}
\usepackage{graphicx}
\usepackage{txfonts}
\usepackage{aa}
\bibstyle{aa}
\def\sn{\hbox{S/N}}  
  
\def\vsin{\hbox{$v \sin i$}}  
  
\def\kms{\hbox{km\,s$^{-1}$}}  
\def\ms{\hbox{m\,s$^{-1}$}}

\def\em{\it}  
  
\def\degr{\hbox{$^\circ$}}

\def\kis{\hbox{$\chi^2$}}   
\def\kisr{\hbox{$\chi^2_{\rm r}$}}   
\def\drot{\hbox{differential rotation}}

\def\porb{\hbox{$(P_{\rm orb}/P_{\rm rot})$}}

\begin{document}

   \title{A maximum entropy approach to detect close-in giant planets around active stars}
   \titlerunning{Maximum entropy detection of planets}

   \author{
          P. Petit
          \inst{1,2}  
          \and
          J.-F. Donati
          \inst{1,2}
          \and
          E. H\'ebrard
          \inst{1,2}
          \and
          J. Morin
          \inst{3}
          \and
          C.P. Folsom
          \inst{4,5}
          \and
          T. B\"ohm
          \inst{1,2}
          \and
          I. Boisse
          \inst{6}
          \and
          S. Borgniet
          \inst{4,5}
          \and
          \\
          J. Bouvier
          \inst{4,5}
             \and
          X. Delfosse
          \inst{4,5}
          \and
          G. Hussain
          \inst{1,2,7}
          \and
          S.V. Jeffers
          \inst{8}
          \and
          S.C. Marsden
          \inst{9}
          \and
          J.R. Barnes
          \inst{10}          
                    }

   \offprints{P. Petit}

   \institute{
   Universit\'e de Toulouse, UPS-OMP, Institut de Recherche en Astrophysique et Plan\'etologie, Toulouse, France \\ 
   \email{ppetit@irap.omp.eu}
\and
CNRS, Institut de Recherche en Astrophysique et Plan\'etologie, 14 Avenue Edouard Belin, F-31400 Toulouse, France
\and
LUPM-UMR 5299, CNRS \& Universit\'e Montpellier II, Place Eug\`ene Bataillon, F-34095 Montpellier Cedex 05, France 
\and
Univ. Grenoble Alpes, IPAG, F-38000 Grenoble, France
\and
CNRS, IPAG, F-38000 Grenoble, France
\and
Universit\'e Aix-Marseille / CNRS-INSU, LAM / UMR 7326, 13388 Marseille, FRANCE
\and
European Southern Observatory, Karl-Schwarzschild-Str. 2, 85748 Garching bei MŸnchen, Germany
\and
Institut f\"ur Astrophysik, Georg-August-Universit\"at G\"ottingen, Friedrich-Hund-Platz 1, 37077 G\"ottingen, Germany            
\and
Computational Engineering and Science Research Centre, University of Southern Queensland, Toowoomba, 4350, Australia
\and
Centre for Astrophysics Research, University of Hertfordshire, College Lane, Hatfield, Herts AL10 9AB, UK
}

   %\date{Received ??; accepted ??}
 
  \abstract
  % context heading (optional)
  % {} leave it empty if necessary  
   {The high spot coverage of young active stars is responsible for distortions of spectral lines that hamper the detection of close-in planets through radial velocity methods.}
  % aims heading (mandatory)
   {We aim to progress towards more efficient exoplanet detection around active stars by optimizing the use of Doppler imaging in radial velocity measurements.}
  % methods heading (mandatory)
   {We propose a simple method to simultaneously extract a brightness map and a set of orbital parameters through a tomographic inversion technique derived from classical Doppler mapping. Based on the maximum entropy principle, the underlying idea is to determine the set of orbital parameters that minimizes the information content of the resulting Doppler map. We carry out a set of numerical simulations to perform a preliminary assessment of the robustness of our method, using an actual Doppler map of the very active star HR~1099 to produce a realistic synthetic data set for various sets of orbital parameters of a single planet in a circular orbit.
   }
  % results heading (mandatory)
   {Using a simulated time series of 50 line profiles affected by a peak-to-peak activity jitter of 2.5~\kms, in most cases we are able to recover the radial velocity amplitude, orbital phase, and orbital period of an artificial planet down to a radial velocity semi-amplitude of the order of the radial velocity scatter due to the photon noise alone (about 50~\ms\ in our case). One noticeable exception occurs when the planetary orbit is close to co-rotation, in which case significant biases are observed in the reconstructed radial velocity amplitude, while the orbital period and phase remain robustly recovered.}
  % conclusions heading (optional), leave it empty if necessary 
   {The present method constitutes a very simple way to extract orbital parameters from heavily distorted line profiles of active stars, when more classical radial velocity detection methods generally fail. It is easily adaptable to most existing Doppler imaging codes, paving the way towards a systematic search for close-in planets orbiting young, rapidly-rotating stars.}

   \keywords{planets and satellites: detection -- stars: imaging -- stars: rotation -- stars: late-type -- stars: planetary systems}

   \maketitle

\section{Introduction}

The numerous detections of exoplanets reported in the last two decades highlight the relatively small population of planets discovered around young stars. These lack of detections, because of the limitations of existing techniques, are especially obvious in studies based on  radial velocity (RV) analysis, with a very low number of planets reported around stars younger than about 300~myr \citep{setiawan07,borgniet14}. The enhanced activity of young Sun-like stars, linked to their fast rotation rate \citep{gallet13}, is the main barrier to RV measurements \citep{saar97, queloz01}. Since the heavy spot coverage of these young objects generates a RV jitter that can hide most planetary signatures, filtering the activity noise is a prerequisite to progress towards a more accurate search for planets orbiting young stars. Our main motivation is to better constrain the formation mechanisms of planetary systems by comparing the observed prevalence of hot jupiters versus model predictions \citep{mordasini09}, as well as to get a more comprehensive view of the evolution of accretion disks and debris disks in the presence of planets \citep{ida05}. 

A number of studies have been dedicated to progress towards more efficient filtering of activity jitter in RV curves (e.g. \citealt{reiners10, boisse12, dumusque14, jeffers14, hebrard14}). For fast-rotating stars,  Doppler imaging (DI) maps can be used to perform a RV filtering taking  a complex distribution of active regions over the stellar surface into account \citep{donati14}. We propose  a different approach, which does not use the two-step method of previous studies in which Doppler mapping is performed prior to the RV analysis. Instead, we implement the recovery of orbital parameters in the DI inversion procedure itself. We therefore do not operate an explicit filtering of the RV curve (we actually do not extract any RVs from our data), but simultaneously get a brightness map and a set of orbital parameters in the output of a slightly modified DI model. Our method shares many similarities with a technique previously implemented to measure stellar rotation and differential rotation \citep{donati00, petit02}, which proved to behave very robustly in a number of practical cases \citep{donati03b, barnes05, petit08, petit10}.

In the rest of the paper, we first present our set of simulated data, detail our maximum entropy approach to extract the orbital elements of an artificial planet, and then present some preliminary tests of the method for various sets of orbital parameters. We finally list our conclusions and mention future tests that need to be performed to fully assess the efficiency of this technique.     

\section{Simulated data}

\begin{figure}
\centering
\includegraphics[width=9cm]{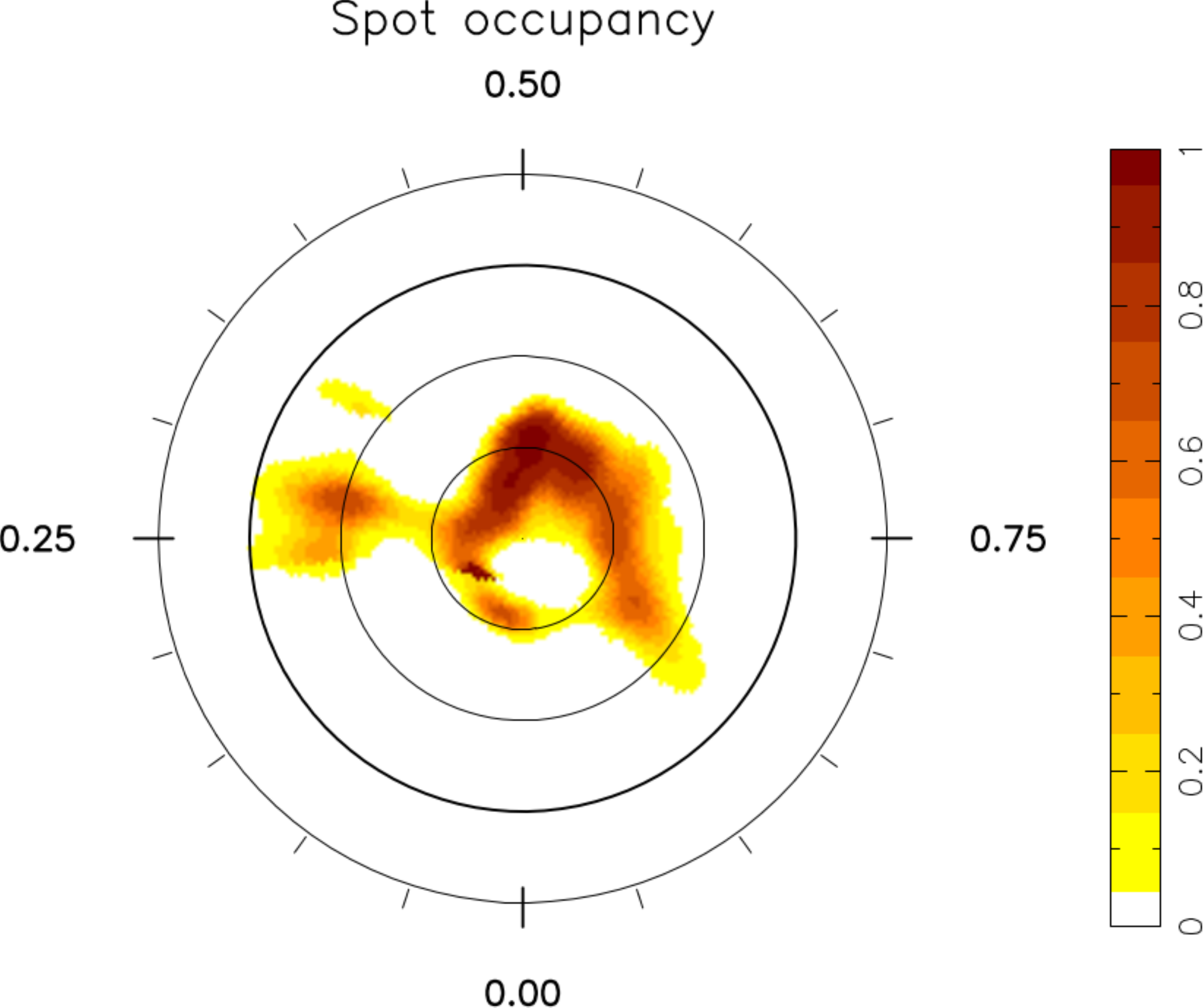}
\caption{Spot occupancy on the artificial star, in flattened polar view, from observations of HR~1099 secured in 1998 January. Concentric circles represent, starting from the centre, parallels of latitude +60, +30, 0, and -30\degr. Radial ticks around the star illustrate the rotational phases, with steps of 0.05 rotation cycle, following \cite{donati03}.}
\label{fig:star}
\end{figure}

The spotted surface of the artificial star considered here is based on a Doppler map of the very active star HR~1099, reproduced on a sphere divided into a grid of 20,000 rectangular pixels of approximately equal area (Fig. \ref{fig:star}). This map was initially reconstructed from data collected at the Anglo-Australian Telescope in 1998 January \citep{donati03} and features a number of large spots distributed from the equator to the observed pole. With a total spot coverage of about six per cent, this specific brightness distribution is a good illustration of the high spot coverage of cool, rapidly rotating stars (e.g. \citealt{donati03, marsden06, marsden11, waite15}). 

\begin{figure}
\centering
\includegraphics[width=6.cm]{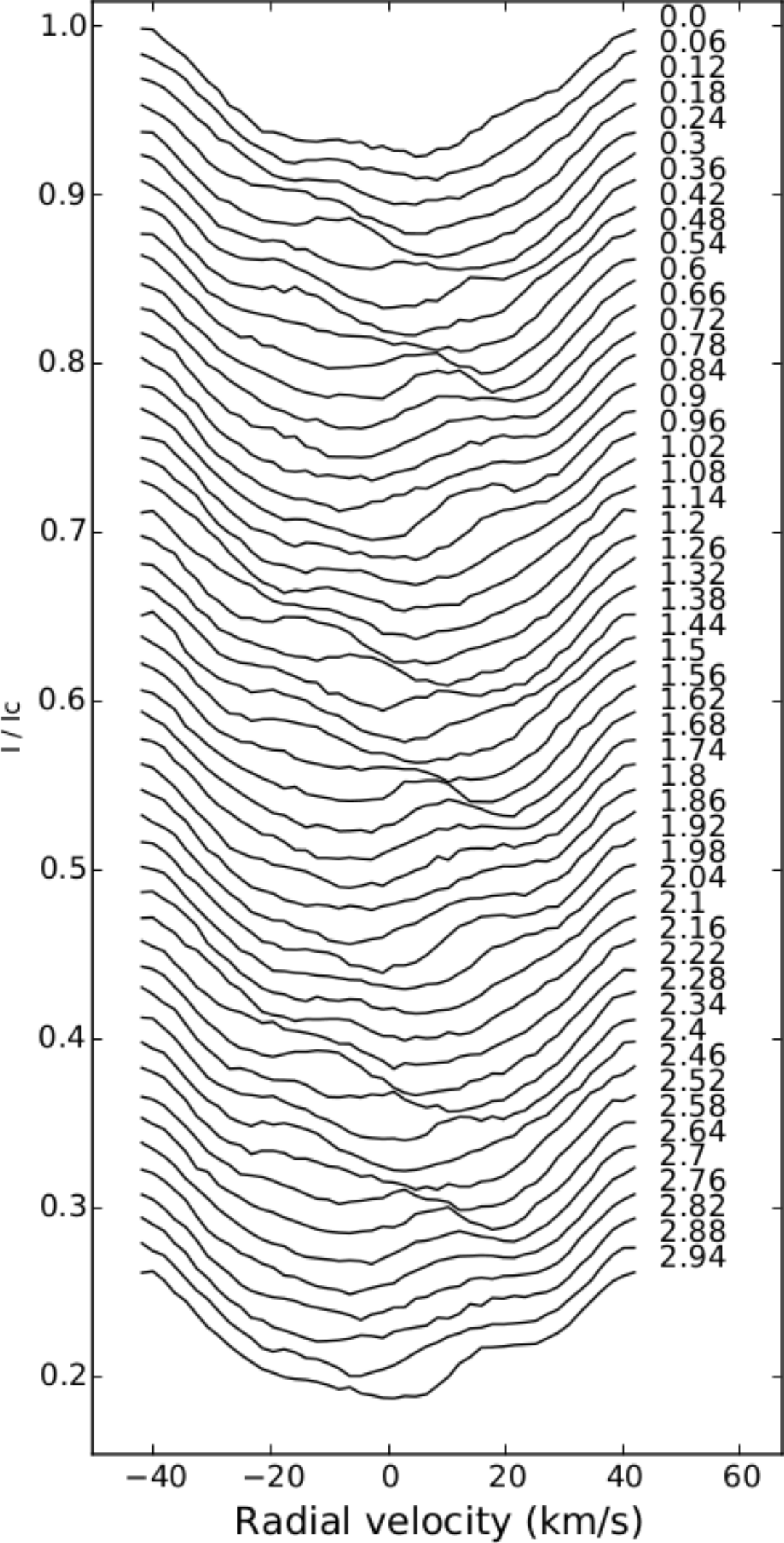}
\caption{Simulated line profiles (normalized to the continuum level) generated from the artificial star of Fig. \ref{fig:star}, in the absence of a planetary companion. Successive profiles are vertically shifted for display clarity and the rotational phase of simulated observation is given in the right side of the plot (with the integer part indicating the rotation cycle number).}
\label{fig:profiles}
\end{figure}

We use this spot distribution to produce a set of synthetic cross-correlation profiles, using the set of tools previously employed by \cite{petit02}. We assume here that the projected rotational velocity (\vsin) of the star is equal to 40~\kms, and that its inclination angle is equal to 38\degr\ (following the values of \citealt{donati03}). The unspotted surface temperature is set to 5,500~K, and the spot temperature is equal to 3,500~K. We assume a spectral resolving power $R=65,000$ (with a Gaussian instrumental profile), and compute line profiles projected on velocity bins of 1.8~\kms for a total of 50 rotational phases evenly spread over three consecutive stellar rotation cycles (Fig. \ref{fig:profiles}). We finally simulate photon noise by adding a Gaussian noise to the line profiles, resulting in a  \sn\ (signal-to-noise ratio) of 1,000 per velocity bin. 

The resulting line profiles are consistent with typical observations of rapidly-rotating, solar-type stars, and,  in particular, young Suns. The \sn\ is at a level commonly reached in a number of published DI studies (e.g. \citealt{donati03}), especially when cross-correlation pseudo-line profiles are modelled instead of individual spectral lines \citep{donati97}. In this case, \sn=1,000 in the cross-correlation profile implies a \sn\ about 30 times lower in the initial spectrum. Gathering 50 rotational phases in a single run and in a time span sufficiently short to prevent any significant spot evolution is demanding but not unrealistic, in particular, for stars with rotation periods of a few days, displaying significant phase evolution within a single observing night. We stress however that the regularity of the phase coverage proposed here is unlikely to be reached from the ground because of observation gaps during the daytime. The  impact of these daytime observations gaps can be at least partly reduced in the case of multi-site campaigns (e.g. \citealt{petit04}). 

\begin{figure}
\centering
\includegraphics[width=9cm]{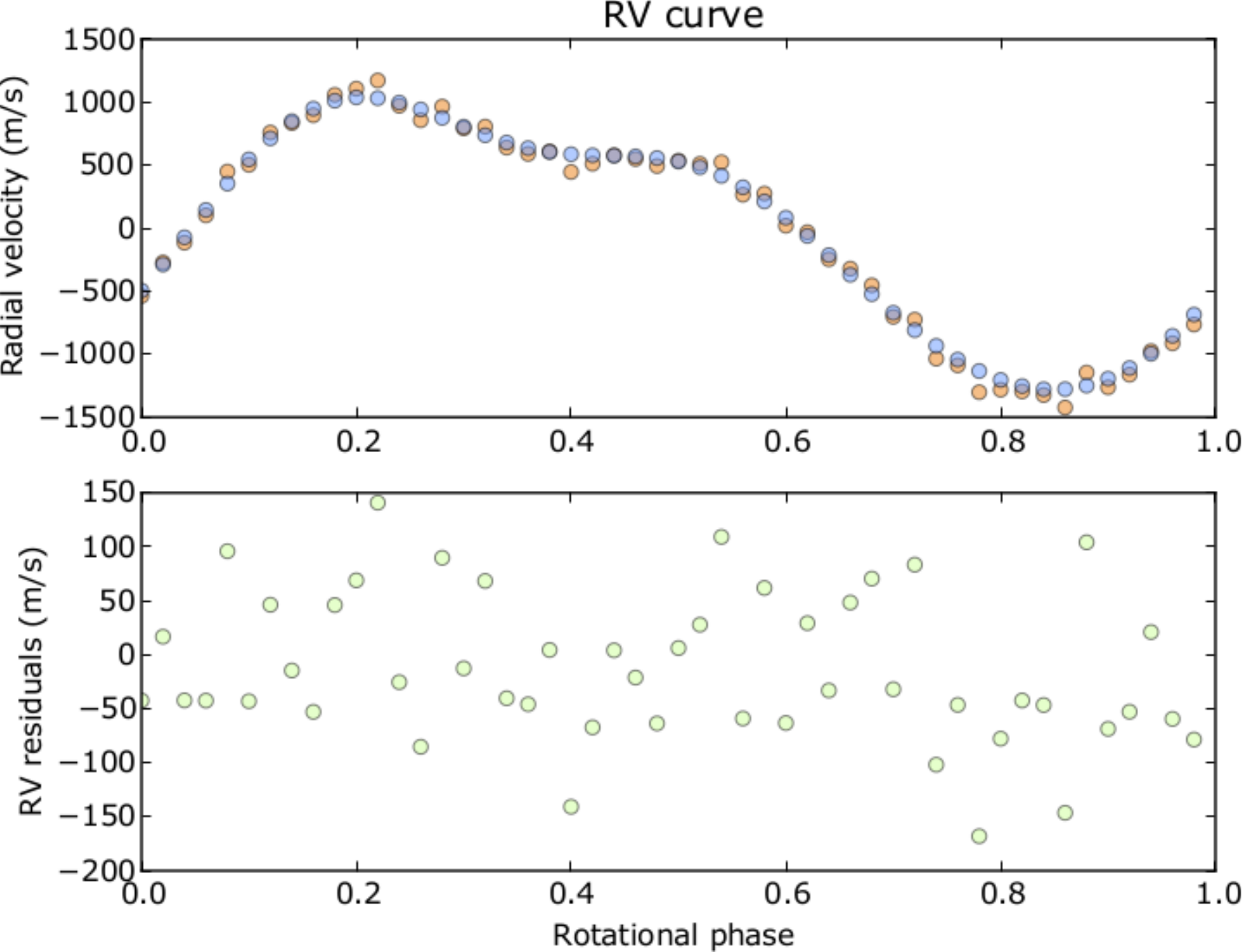}
\caption{Radial velocity time series extracted from our set of simulated cross-correlation profiles using the first moment method (top panel, orange dots), plotted together with radial velocities produced by the Doppler imaging code (blue dots). This model does not include any orbiting planet. The bottom panel shows the RV residuals.}
\label{fig:rvcurve}
\end{figure}

The $rms$ RV jitter for this fake data set is equal to 0.8 \kms, as estimated from the standard deviation of the first moments of the line profiles (about 2.5~\kms\ peak to peak, see Fig. \ref{fig:rvcurve}). We note that the Gaussian noise alone is responsible for a RV $rms$ scatter of 66~\ms\ only (this value being a function of the adopted \sn\ and velocity bin size), so that the standard deviation of RVs is dominated by the simulated spot activity. 

We finally alter the line profiles to incorporate RV shifts due to the presence of a close-in planet, assuming  that the instrument benefits from perfect RV stability. For simplicity, we assume  here a single planetary companion, following a circular orbit. The orbit and stellar equator share the same plane. The time-dependent RV shift induced by the companion is defined by the ratio of orbital over (stellar) rotation periods ($P_{\rm orb}/P_{\rm rot}$), its semi-amplitude ($K$), and its phase delay ($\phi$) compared to stellar rotation. We define the phase delay so that it is equal to zero if the first observed stellar rotational phase (also set to zero) is a time of maximal planetary-induced RV shift. 

\section{Simultaneous maximum entropy inversion of the brightness map and orbital parameters}

\subsection{Method outline}

\begin{figure}
\centering
\includegraphics[width=8cm]{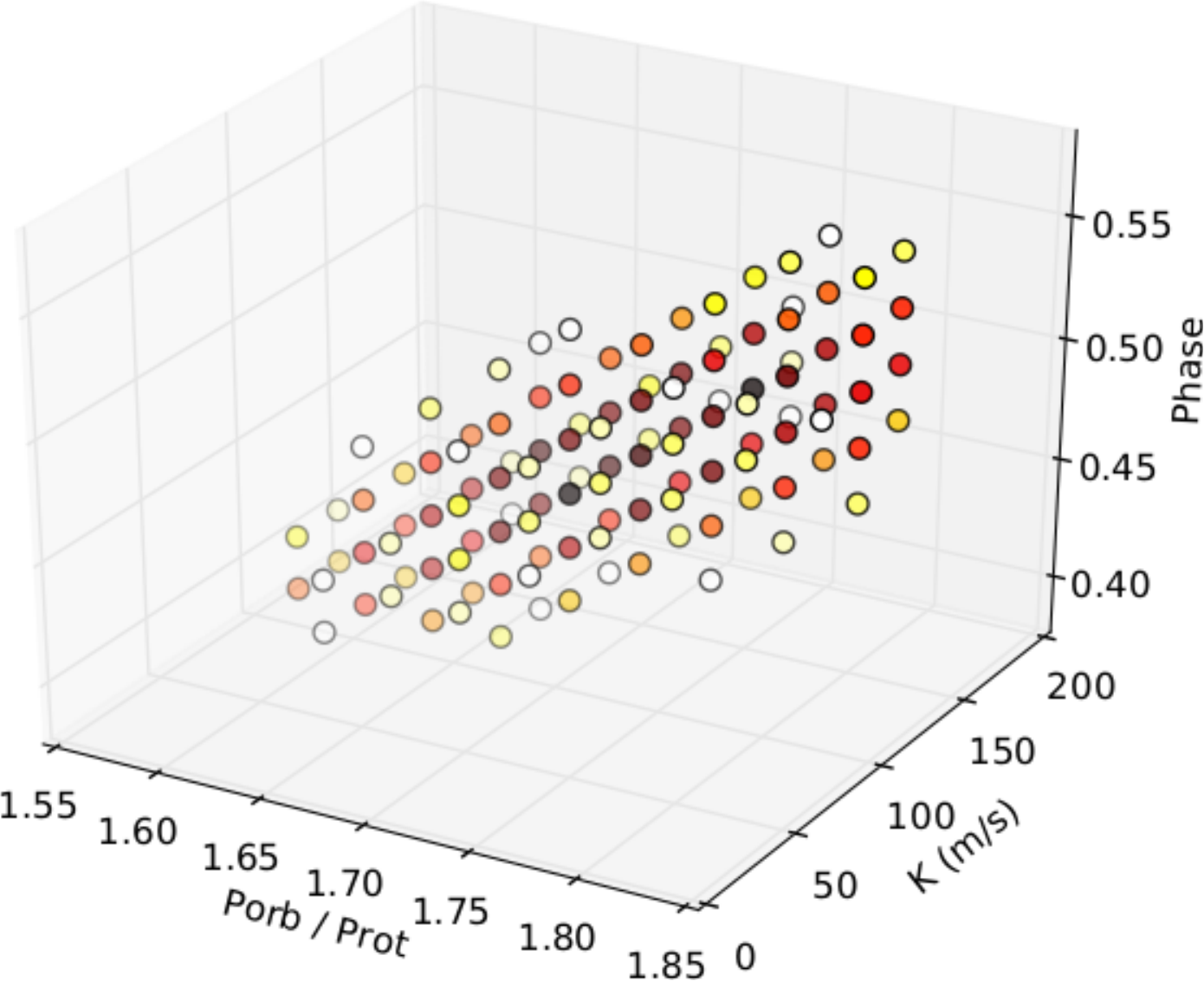}
\includegraphics[width=8cm]{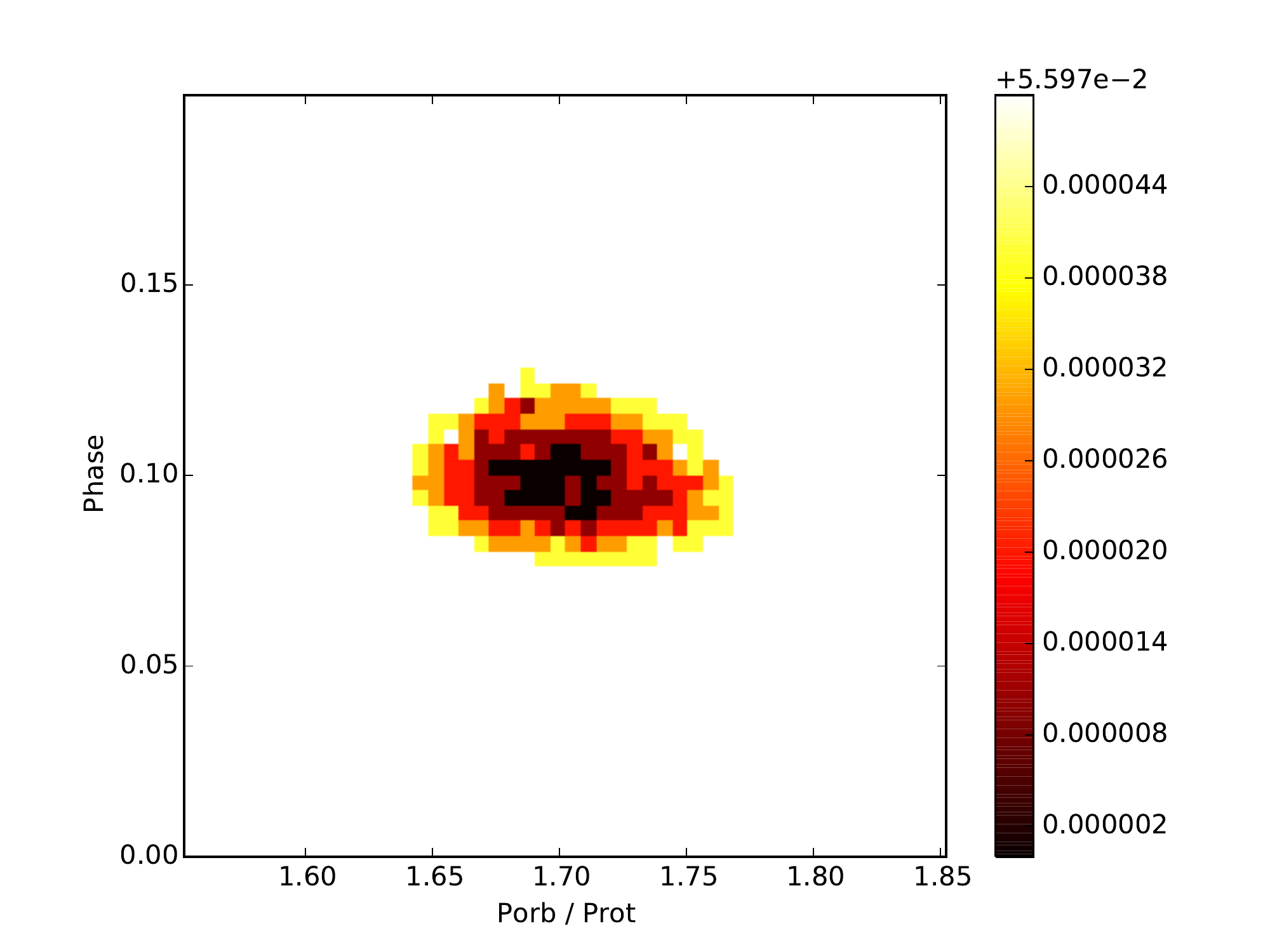}
\includegraphics[width=8cm]{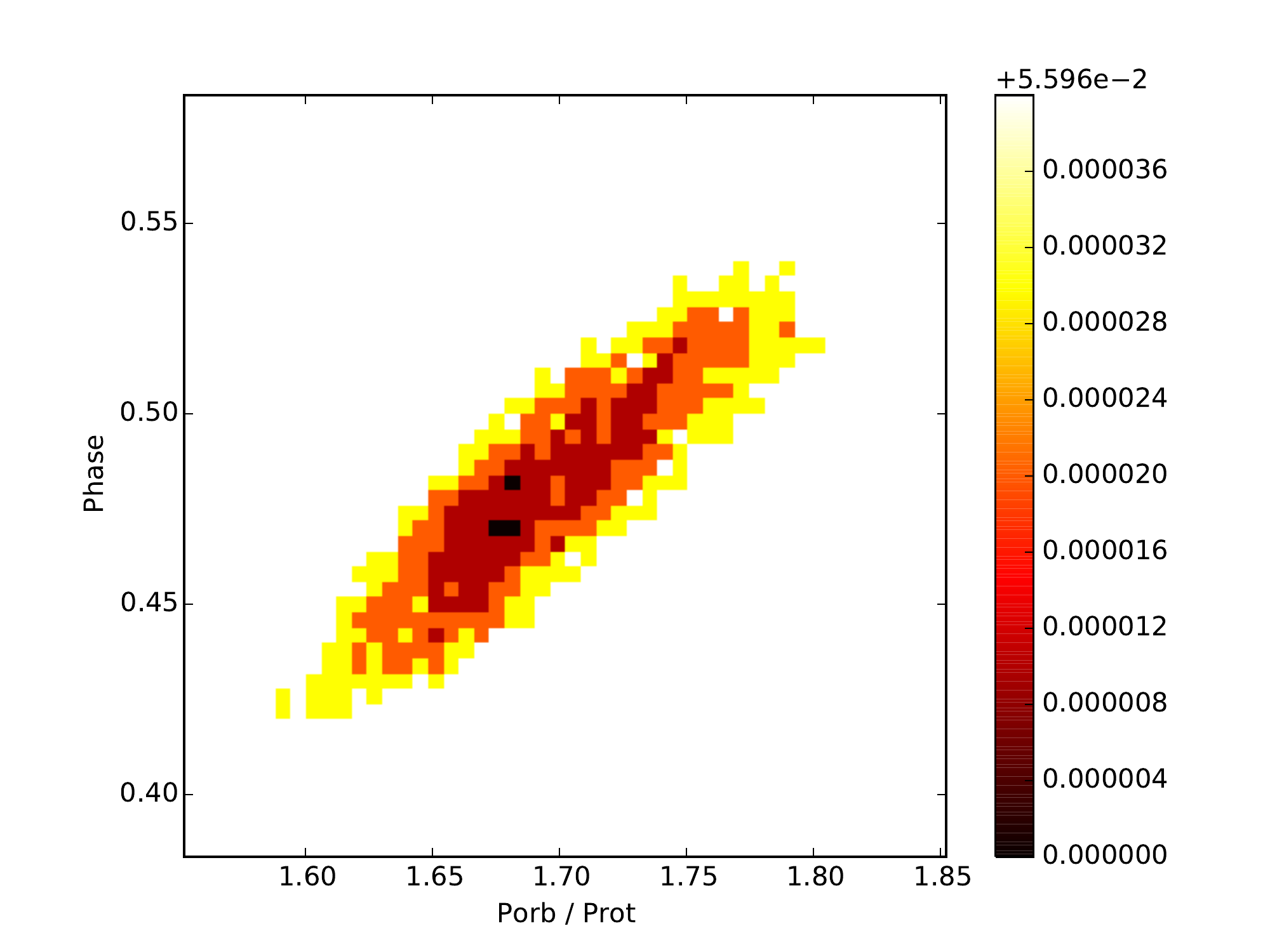}
\caption{Top panel: example 3D entropy (spottedness) landscape, obtained from 1,000 DI models with different values of $K$, $P_{\rm orb}/P_{\rm rot}$, and $\phi$. The simulated parameter values were $K=0.1$~\kms\, $P_{\rm orb}/P_{\rm rot}=1.7$ and $\phi=0.5$. Middle and bottom panels: example 2D slices in two orthogonal planes, extracted from the 3D orbital parameter space. Each slice is obtained from a set of 2,500 Doppler images. In all three panels, the colour scale shows the fractional spot coverage of the resulting Doppler maps, and the location of minimal spottedness shows the reconstructed set of orbital parameters.}
\label{fig:landscape}
\end{figure}

We use here a maximum entropy approach to simultaneously reconstruct the stellar brightness map and orbital parameters from the set of simulated observations. With this strategy, orbital parameters are directly included in the DI model, similarly to relevant stellar parameters (e.g. stellar inclination or \vsin). Following the very basic principle of maximum entropy image reconstruction, the most likely value for the triplet of orbital parameters is the value that produces a brightness map with the lowest information content, i.e. the reconstructed map featuring the lowest spot coverage. 

In practice, we adapt the method successfully implemented by \cite{petit02} to recover the surface \drot\ of active stars, by computing a grid of Doppler images spanning a range of values of $K$, $P_{\rm orb}/P_{\rm rot}$, and $\phi$. This method is directly inspired from the "entropy landscape" method of \cite{donati00} and \cite{watson01}, although transposed to a 3D parameter space. We reconstruct the brightness maps using the DI code of \cite{donaticameron97}, which implements the two-component model of \cite{cameron92}. This model is based on the assumption that each pixel of the reconstructed stellar surface contains a fraction $f$ of unspotted photosphere and a fraction $1 - f$ of dark spot. From the same simulated data set (i.e. for the same simulated value of $K$, $P_{\rm orb}/P_{\rm rot}$ and $\phi$), we reconstruct a set of 1,000 to 3,375 Doppler maps, assuming different values of the ($K$, $P_{\rm orb}/P_{\rm rot}$,$\phi$) triplet in the inversion process. We derive from each reconstructed map a value of the fractional spot coverage and therefore obtain a 3D spottedness landscape in the ($K$, $P_{\rm orb}/P_{\rm rot}$,$\phi$) space (Fig. \ref{fig:landscape}). To determine the reconstructed value of ($K$, $P_{\rm orb}/P_{\rm rot}$,$\phi$), we adjust a 3D paraboloid in the ($K$, $P_{\rm orb}/P_{\rm rot}$,$\phi$) space, limiting the fit to a few tens of points around the point of minimal spottedness. By locating the point of minimal information content of the Doppler map, the paraboloid fit gives the most likely output value for ($K$, $P_{\rm orb}/P_{\rm rot}$,$\phi$). 

Similar to the method adopted by \cite{petit02}, error bars shown in Fig. \ref{fig:amp}, \ref{fig:porb}, and \ref{fig:phase} are derived from the paraboloid fit by determining the projection on the three axis of the $\Delta \chi^2=1$ surface surrounding the $\kis$ minimum, following the prescription of \cite{press92}. Since the reconstructed orbital parameters are computed from DI models, their estimate benefits from an efficient noise reduction technique (the maximum entropy method), which  tends to decrease the uncertainties in their reconstructed values beyond the capabilities of a simple \kis\ minimization approach. In this sense, using \kis\ values alone to derive statistical uncertainties, as we do here, implies that all error bars discussed in the following sections are likely over-estimated.

We do not perform  an explicit filtering of activity jitter in line profiles, as done by \cite{donati14}. Instead, we obtain the orbital parameters as part of the image reconstruction itself, although the reconstructed line profiles produced by the DI code, in principle, can be used to perform a jitter filtering of observational data to double check the orbital parameters identified with our method (Fig \ref{fig:rvcurve}). We emphasize that the rough atmospheric model used in the DI code is the same as that used to produce our fake data, so that we implicitly assume here that the DI atmospheric model is perfect. 

\subsection{Test case without a planet}

\begin{figure}
\centering
\includegraphics[width=8cm]{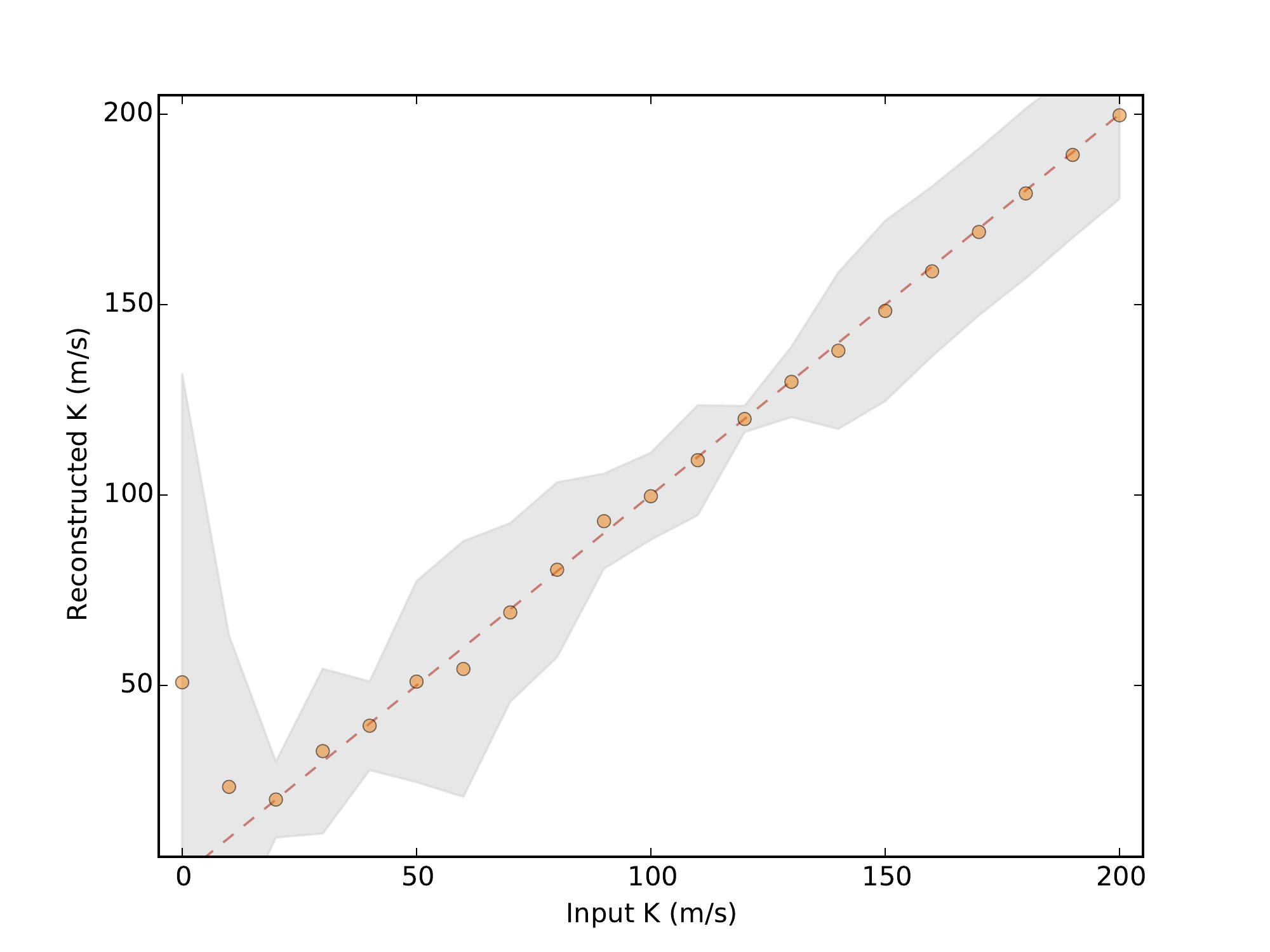}
\includegraphics[width=8cm]{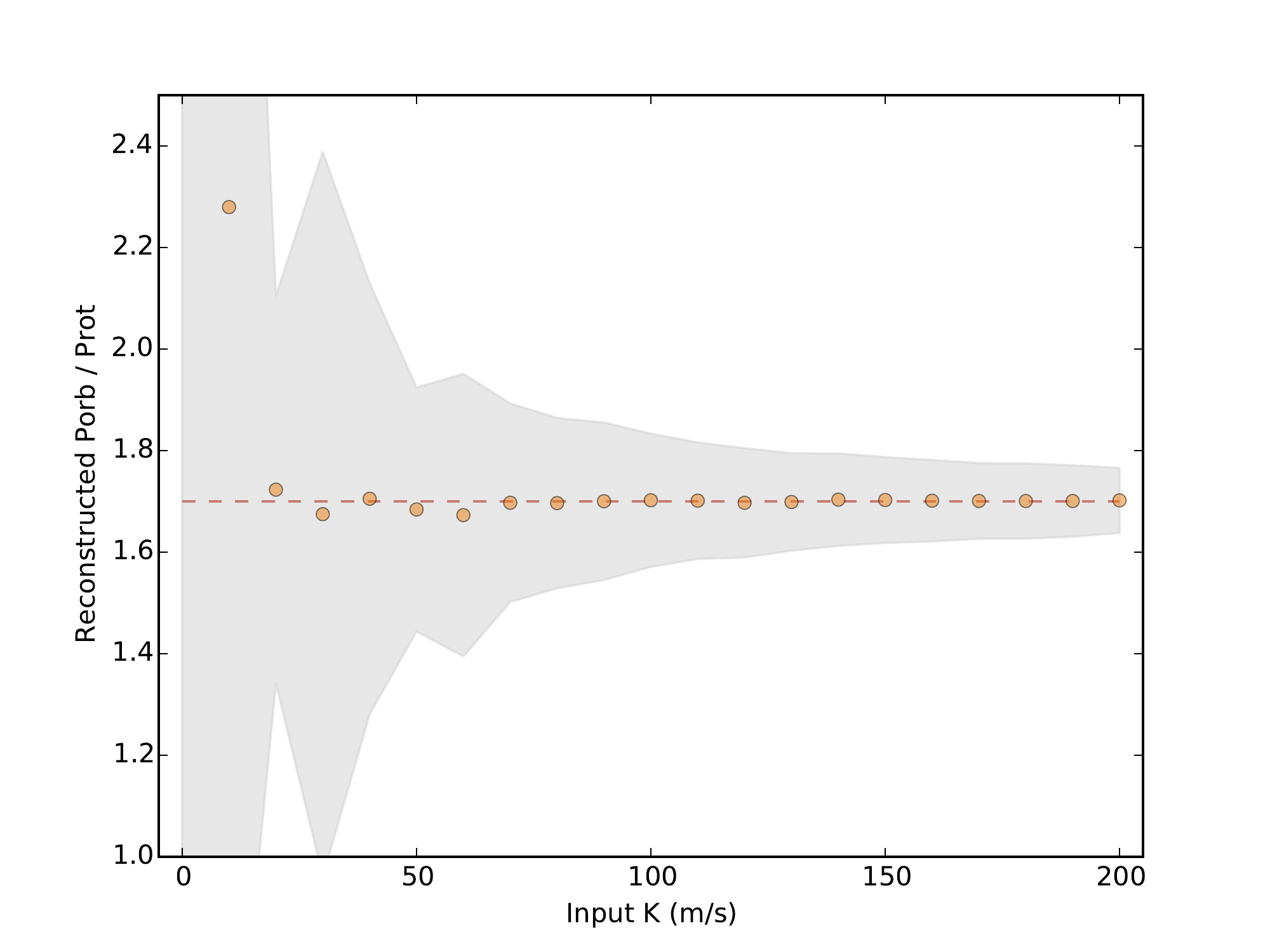}
\includegraphics[width=8cm]{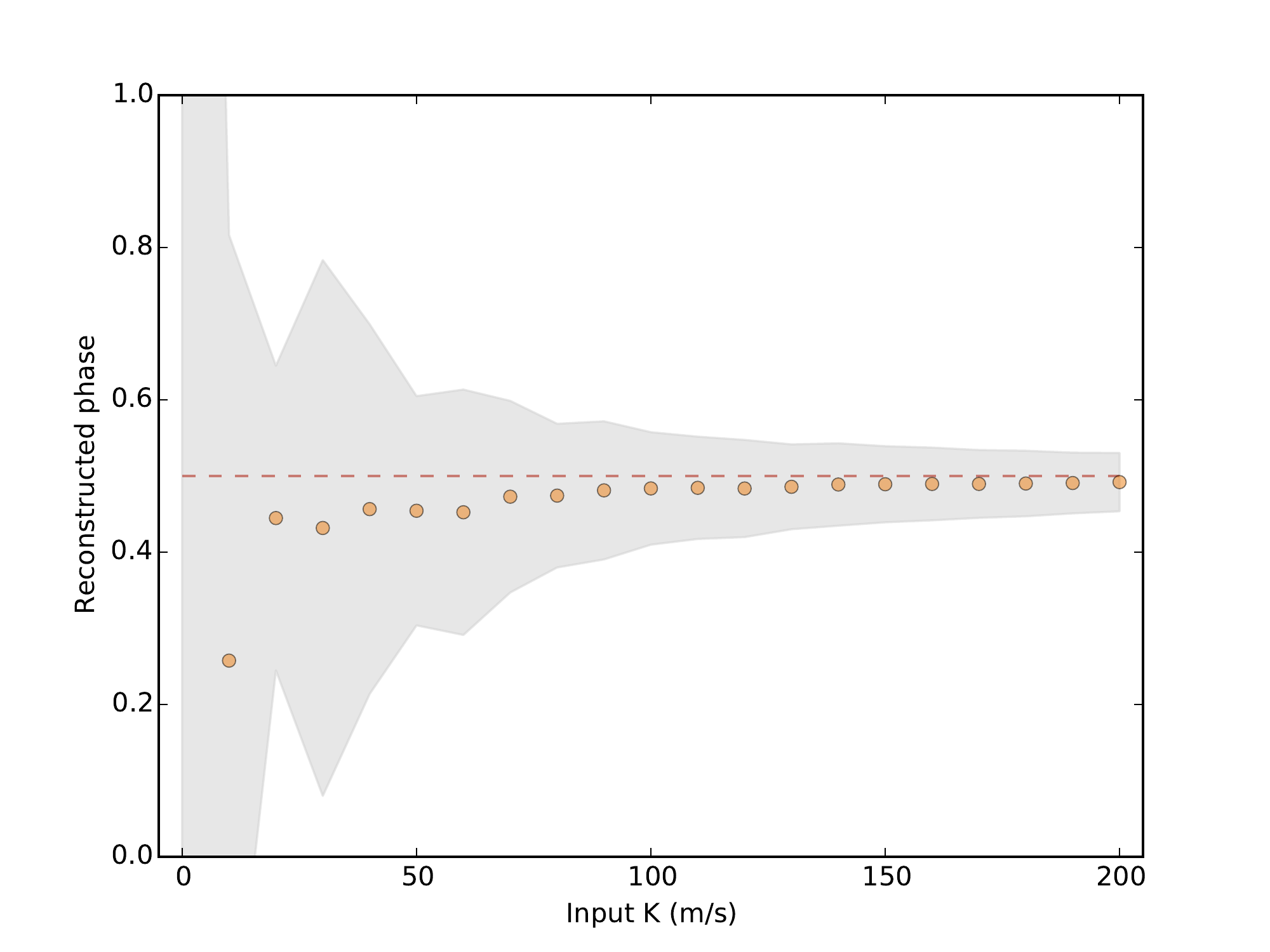}
\caption{Simulations for a radial velocity amplitude $K$ varying from 0 to 200~\ms. The orbital period is fixed and equal to 1.7 times the rotational period, as well as $\phi$, which is set to 0.5. The dashed lines indicates the input values, while circles show the reconstructed parameters. The $1\sigma$ error bars derived on individual parameters from the paraboloid fit are plotted as grey areas.}
\label{fig:amp}
\end{figure}

The first test of our technique consisted of simulating a star without any planet ($K$, $P_{\rm orb}/P_{\rm rot}$,$\phi$)=(0,0,0) and checking the outcome of the Doppler inversion. The simulated set of line profiles is shown in Fig. \ref{fig:profiles} and the resulting map (not shown here) is reconstructed with a reduced \kis\ close to unity (\kisr=0.97). The initial spot configuration is correctly recovered, with the location and area of the reconstructed spots consistent with the initial spot configurations. Using the set of line profiles produced by the DI code to remove the activity-induced RV fluctuations from our simulated data, we end up with a  $rms$ residual jitter of 68 \ms, which is very slightly larger than the $rms$ scatter induced by the simulated photon noise alone (bottom panel of Fig. \ref{fig:rvcurve}), showing that our simulated observations were adjusted down to the noise level, with no detectable residuals of the activity jitter in the filtered RV curve. 

The reconstructed orbital parameters are shown as the leftmost points of Fig. \ref{fig:amp}. As expected in this test case, $P_{\rm orb}/P_{\rm rot}$ and $\phi$ turn out to be unconstrained in the entropy space. Their error bars are extremely large, translating the fact that the entropy landscape is mostly flat in these two directions. The reconstructed value of $K$ is equal to $51 \pm 81$ \ms, showing that the modified DI code cannot reconcile this simulated data set with a RV semi-amplitude higher than about $\approx 130$ \ms. The relatively large error bar on the reconstructed $K$ obtained at $K_{\rm in}=0$, compared to higher $K_{\rm in}$ values (see Sect. \ref{sec:kvar}), is a consequence of the correlation between the different parameters of the triplet, as observed in Fig. \ref{fig:landscape} (so that the large error bars on the estimate of $P_{\rm orb}/P_{\rm rot}$ and $\phi$ also tend to enlarge the error bar on $K$).

\subsection{Orbital parameter reconstruction}

The next test consisted of running a series of simulations with different input values of ($K$, $P_{\rm orb}/P_{\rm rot}$, $\phi$) to test the robustness of the method for different orbital configurations.

\subsubsection{Variations of $K$}
\label{sec:kvar}

\begin{figure}
\centering
\includegraphics[width=8cm]{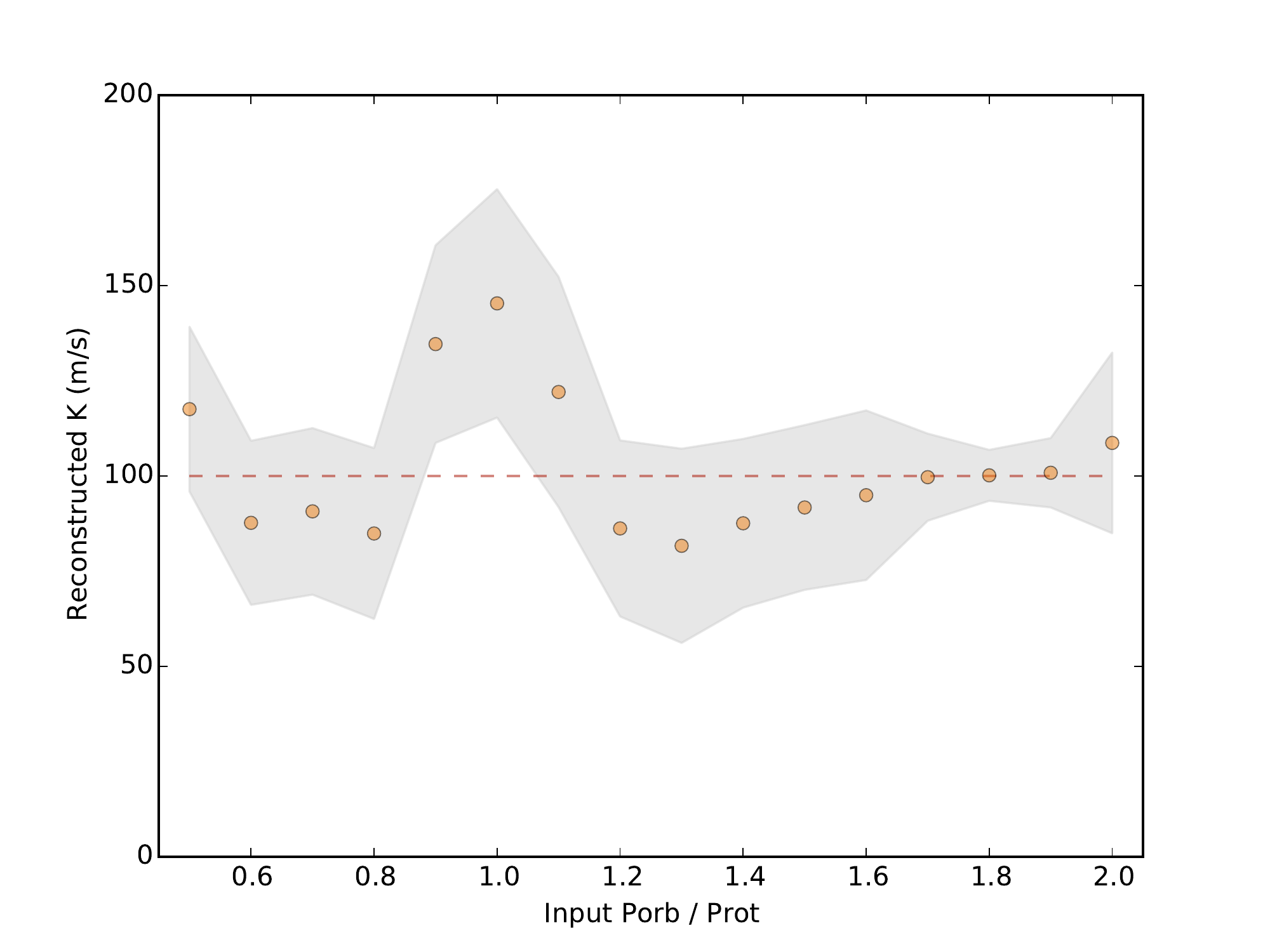}
\includegraphics[width=8cm]{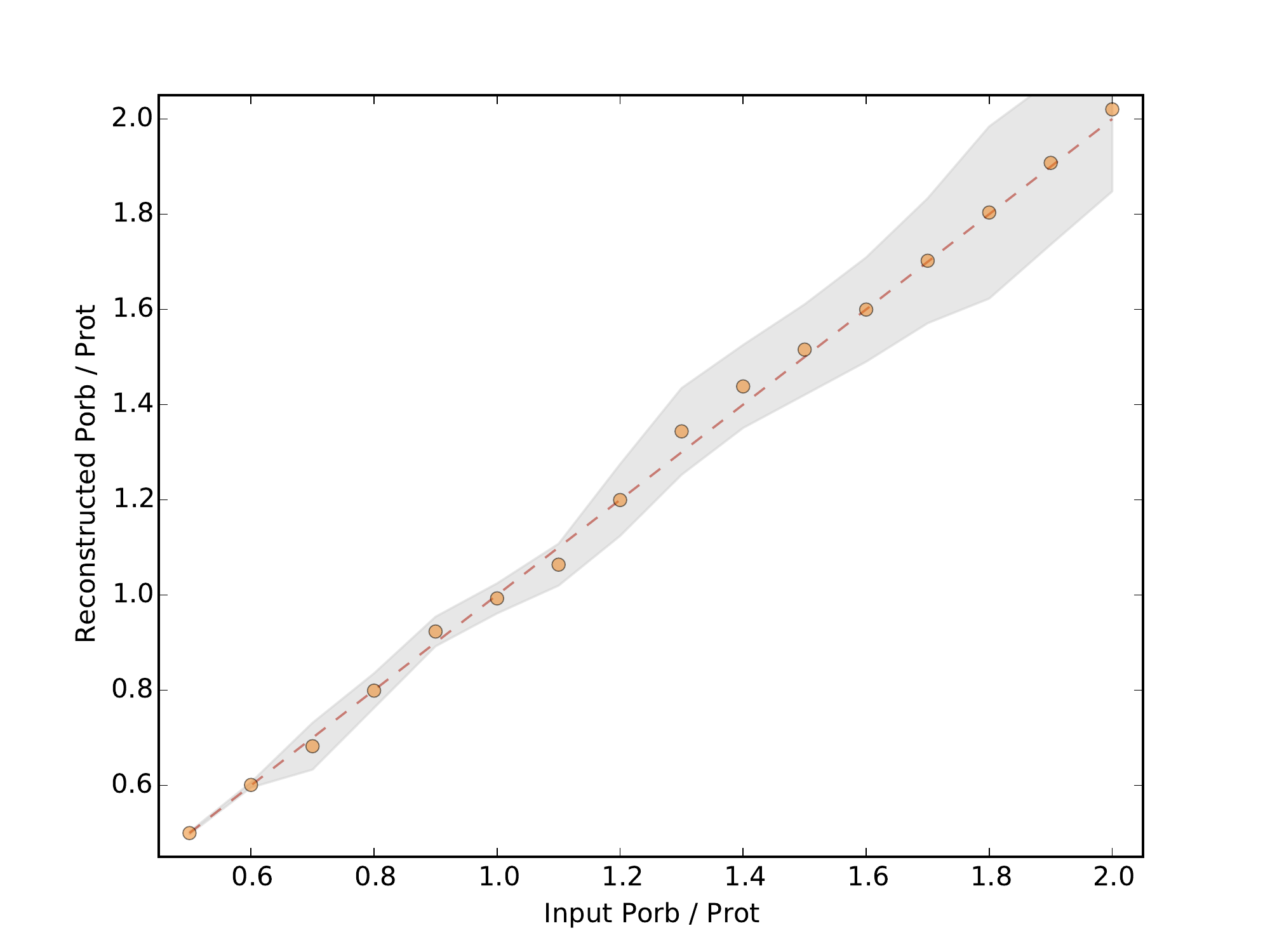}
\includegraphics[width=8cm]{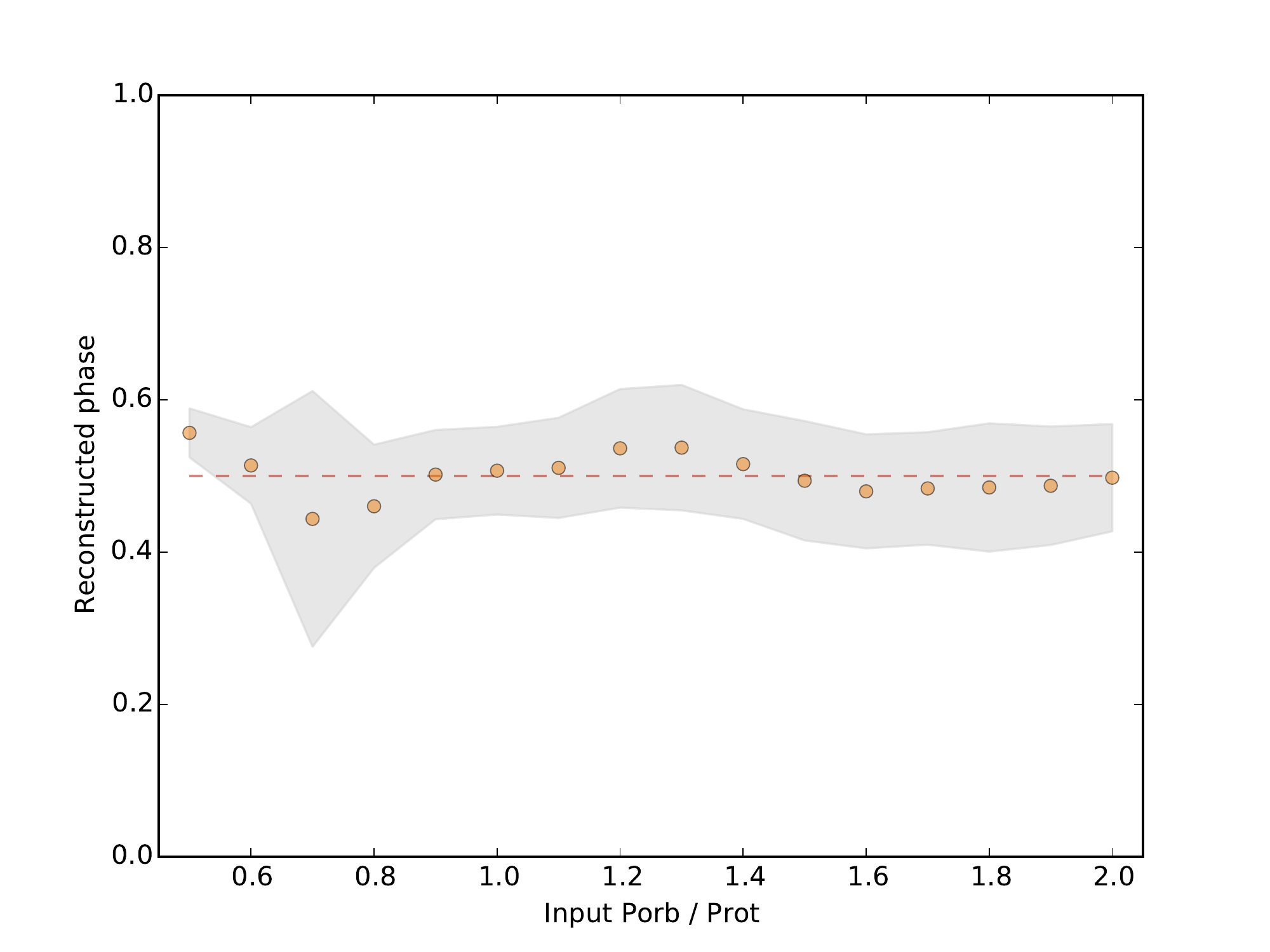}
\caption{Same as Fig. \ref{fig:amp}, except for an orbital period varying from 0.5 to 2 times the stellar rotation period. The radial velocity amplitude $K$ is fixed and equal to 100 \ms and $\phi$, which is set to 0.5.}
\label{fig:porb}
\end{figure}

We first assumed a constant simulated value of  $(P_{\rm orb}/P_{\rm rot})_{\rm in}$ (equal to 1.7), as well as a constant phase delay $\phi_{\rm in}=0.5$. We then varied our simulated $K_{\rm in}$ value from 0 to 200 \ms\ (Fig. \ref{fig:amp}). We observe that the reconstructed values of $K$ are very close to the input values (to within 10 per cent) as long as $K_{\rm in} > 20$~\ms, with typical error bars smaller than about 20~\ms\ (up to 35 \ms\ at $K_{\rm in}=60$~\ms). For even smaller simulated values of $K_{\rm in}$, the reconstructed values display much larger biases towards higher RV amplitudes, with error bars increasing very fast when $K_{\rm in}$ approaches zero.

The reconstructed values of $P_{\rm orb}/P_{\rm rot}$ are also very close to the input values (to within 2$\%$), as long as the input $K_{\rm in}$ is larger than 20~\ms. The error bars on $P_{\rm orb}/P_{\rm rot}$  progressively increase while $K_{\rm in}$ decreases, going from 0.05 at $K_{\rm in} = 200$~\ms\ to 0.6 at  $K_{\rm in} = 30$~\ms, and remain large whenever $K_{\rm in} < 50$~\ms. The reconstruction biases also become much larger for $K_{\rm in} < 20$~\ms

The behaviour of the reconstructed $\phi$ is mostly similar to that of $P_{\rm orb}/P_{\rm rot}$, with more systematic biases at low values of $K_{\rm in}$. 

\subsubsection{Variations of $P_{\rm orb}/P_{\rm rot}$}
\label{sect:porb}

\begin{figure}
\centering
\includegraphics[width=8cm]{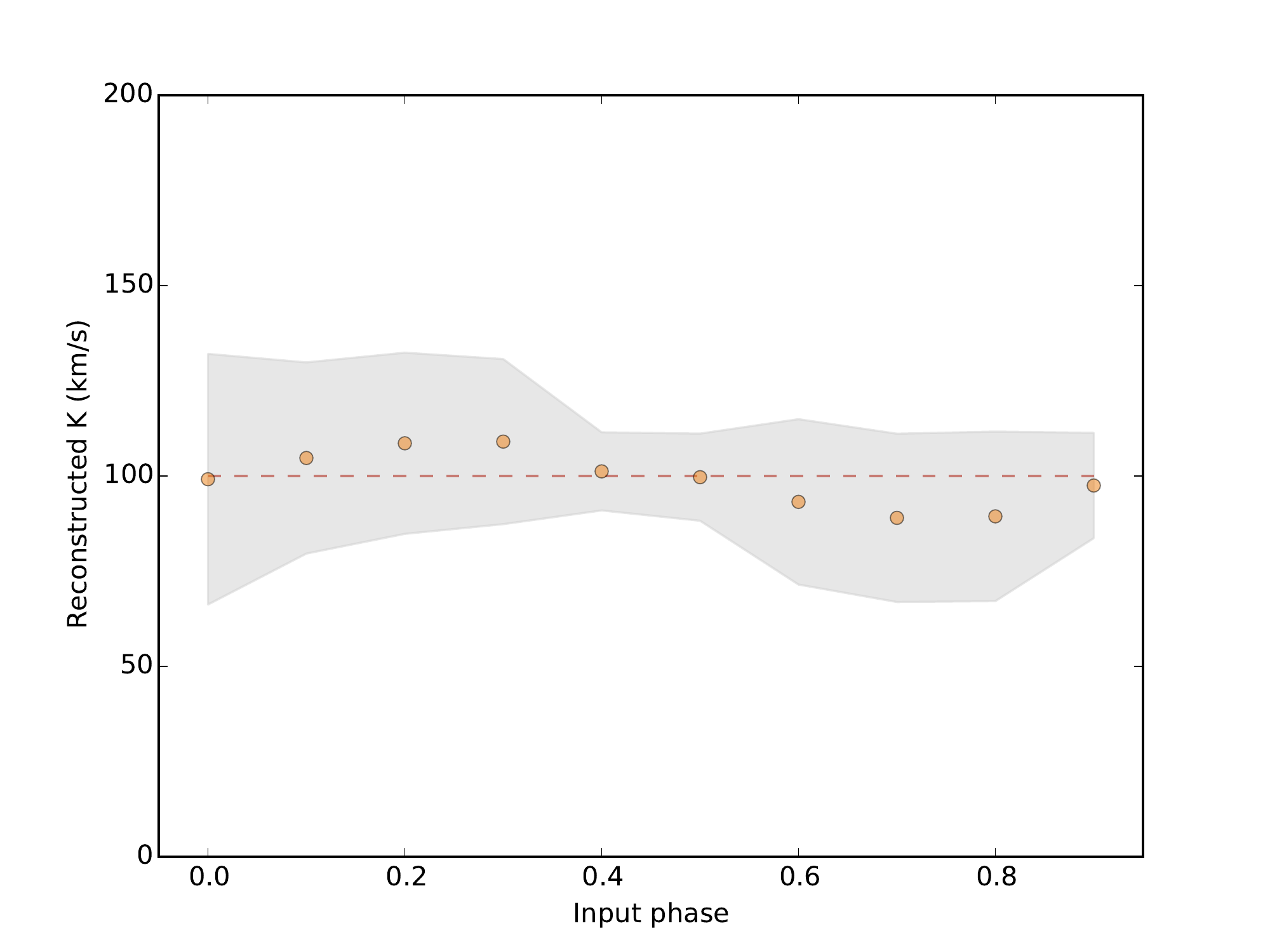}
\includegraphics[width=8cm]{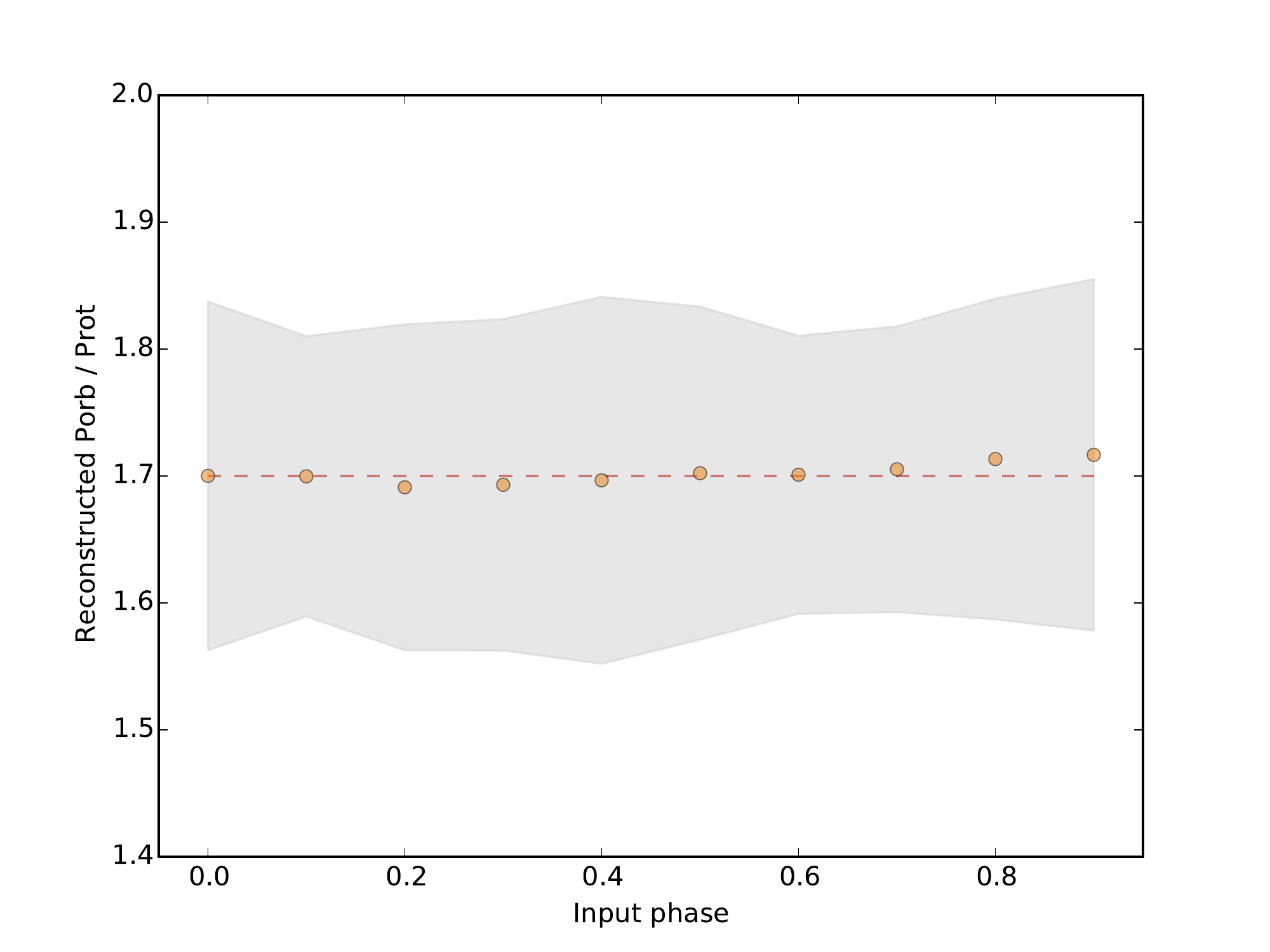}
\includegraphics[width=8cm]{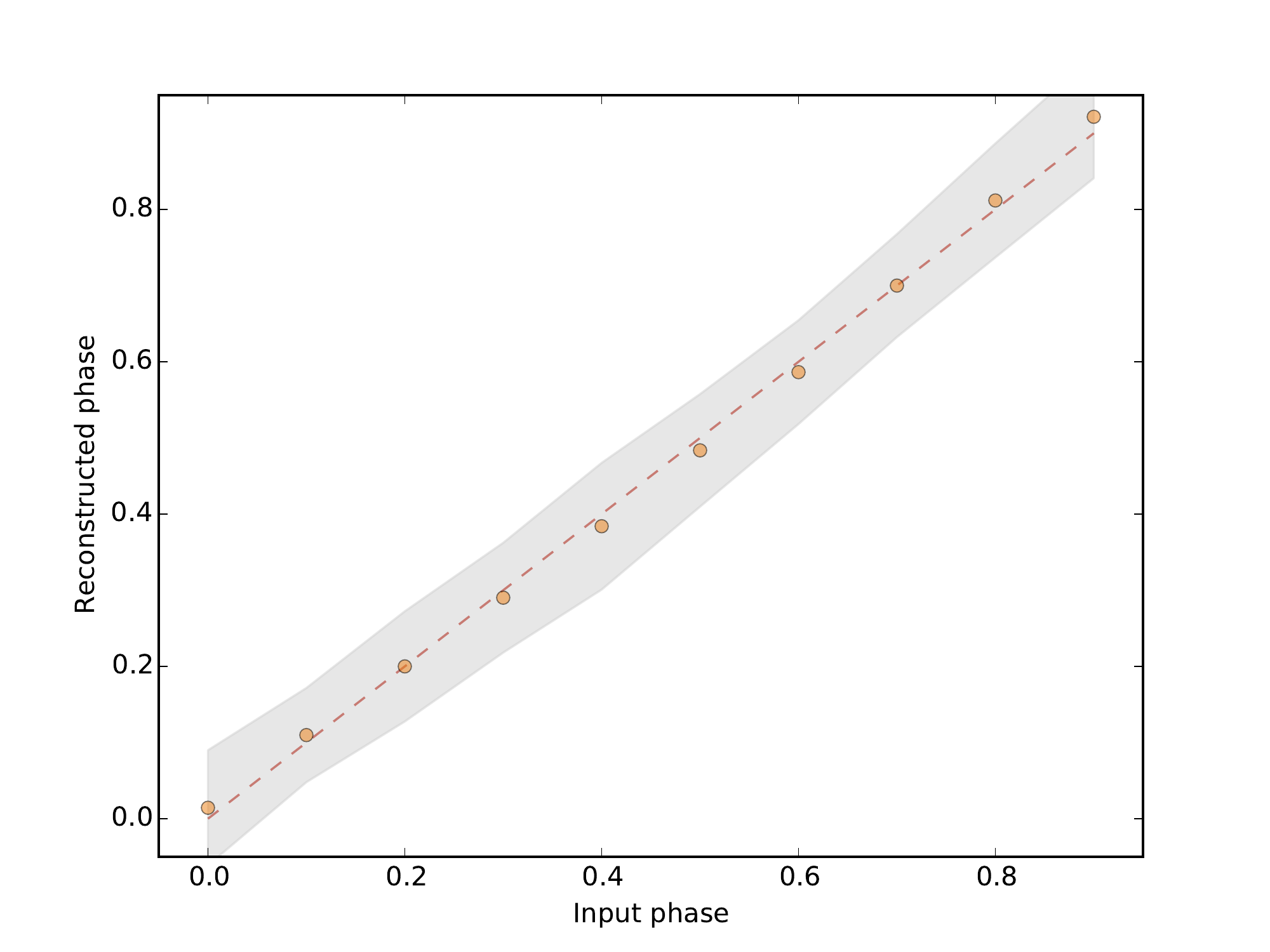}
\caption{Same as Fig. \ref{fig:amp}, except for an orbital phase varying from 0 to 0.9. The radial velocity amplitude $K$ is fixed and equal to 100 \ms and $P_{\rm orb}/P_{\rm rot}$, which is set to 1.7.}
\label{fig:phase}
\end{figure}

We also performed a series of simulations at fixed values of $K_{\rm in}=100$~\ms, and $\phi_{\rm in}=0.5$. Here, $(P_{\rm orb}/P_{\rm rot})_{\rm in}$ was varied from 0.5 to 2 (Fig. \ref{fig:porb}). The reconstructed \porb\ is always close to its input value, without any systematic bias. The error bar on the recovered value of \porb\ is steadily increasing towards larger values of \porb$_{\rm in}$, showing that it is preferable to spread the observations over several orbital periods, rather than getting a denser sampling of one single orbital cycle.

The most striking feature in the reconstructed $K$ values is a clear bias of up to 50\% when $P_{\rm orb} \approx P_{\rm rot}$, although the reconstructed \porb\ and $\phi$ are much less affected by this phenomenon. The poor reconstruction obtained for a planet close to co-rotation, already reported by \cite{donati14}, illustrates that the modified DI code cannot fully distinguish between a spot and a planetary signature as long as the rotational and orbital modulations are not following a different period. As illustrated in Fig. \ref{fig:errors}, adding a different noise pattern to our data (with a same resulting \sn) does not remove this bias. Apart from this zone, biases in the reconstructed $K$ do not exceed 20\%, and get even better at \porb$> 1.6$.

\subsubsection{Variations of $\phi$}

The last test consisted of varying $\phi_{\rm in}$, at fixed values of $K_{\rm in}$ and \porb$_{\rm in}$ (Fig. \ref{fig:phase}). The outcome is an accurate reconstruction of the triplet for all values of $\phi_{\rm in}$, with limited biases for all orbital parameters, suggesting that the $\phi$ parameter is the less critical of the triplet in this reconstruction procedure.  

\section{Impact of uncertainties in reconstruction parameters}

\begin{figure}
\centering
\includegraphics[width=8cm]{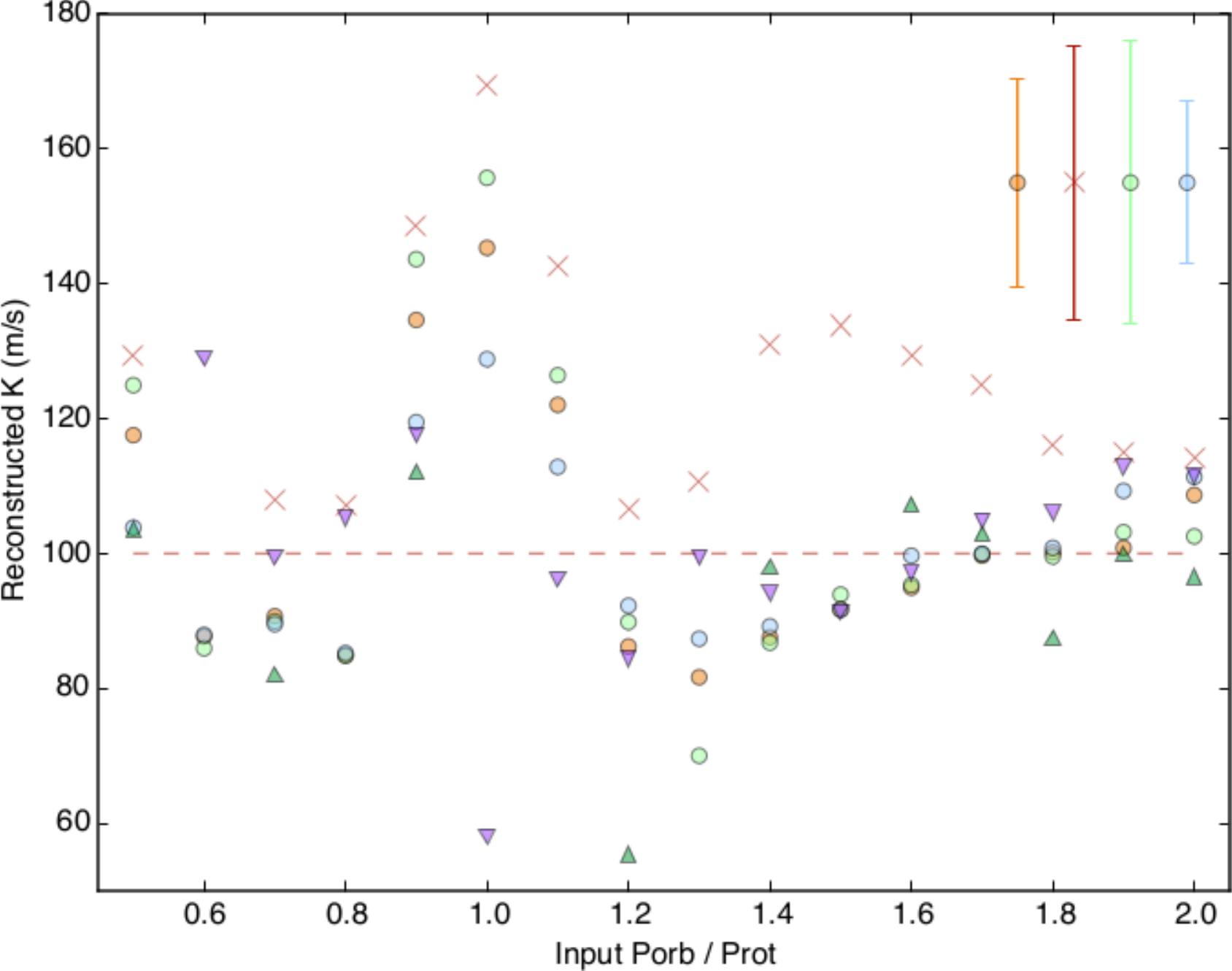}
\caption{Same as the top panel of Fig. \ref{fig:porb}, with errors introduced in the reconstruction parameters. Orange circles again show the points of Fig. \ref{fig:porb}, obtained with $\chi_r^2=0.97$ and $i=38$\degr, $i.e.$ our optimal model.  Green and pale blue circles are obtained by forcing $\chi_r^2=1.0$ and 0.94, respectively, during reconstruction. Violet (resp. green) triangles show the effect of forcing $i=33$\degr\ (resp. $i=43$\degr). Crosses are obtained by adding another noise pattern to our fake data set (using a different seed to generate random numbers, still resulting in \sn$\approx$1,000). Typical error bars are indicated in the top right side of the plot, except for violet and green triangles (affected by much larger statistical error bars that would hamper the plot readability).}
\label{fig:errors}
\end{figure}

In our ideal numerical simulations, setting reconstruction parameters in the DI code is straightforward in  that we  know the parameters used to generate the artificial star and fake time series of observations exactly. This question is notoriously more difficult with real data, and errors in the parameters adopted in the DI procedure are known to generate biases in the reconstruction of stellar brightness maps \citep{unruh95}. We carried out a preliminary investigation of the outcome of wrong values of DI input parameters in the reconstructed triplet of orbital parameters. We based this new series of tests on simulations similar to those discussed in Sect. \ref{sect:porb}, i.e. where the RV amplitude $K$ is fixed and equal to 100 \ms, as well as $\phi$, which is set to 0.5, while $P_{\rm orb}/P_{\rm rot}$ is varied between 0.5 and 2.0. The reconstructed $K$ values obtained in this new series of simulations are shown in Fig. \ref{fig:errors}.

A first test consisted in varying the final \kisr\ values of the models. Instead of the default target value of 0.97, which is set to be the highest \kisr\ value below which the $rms$ of the RV residuals cannot be reduced further, \kisr\ = 1.0 and 0.94 are successively adopted (circles in Fig. \ref{fig:errors}). We observe that reconstruction biases in $K$ values tend to decrease for decreasing \kisr\ values and the associated error bars. However,  the observed changes remain limited compared to other effects. In particular, adopting  \kisr\ = 0.97 again with a different noise pattern in our data (simply obtained with a different seed to generate random numbers, red crosses in Fig. \ref{fig:errors}) roughly leads  to the same amplitude of changes in the biases and error bars. Similar conclusions are reached for the reconstruction of $P_{\rm orb}/P_{\rm rot}$ and $\phi$ (not shown here). Below \kisr\ = 0.94, most of our models did not converge at all.

As a second test, we modified the inclination angle assumed during the inversion process, adopting $i=33$ and 43\degr\ instead of  $i=38$\degr\ (triangles in Fig. \ref{fig:errors}). While some of these models still provided us with reconstructed $K$ values close to the input value, the biases become much larger in a number of cases to the point where several points fall outside of the boundaries of our figure. The obvious inaccuracy in the inclination angle is also highlighted by the best achievable \kisr\ , which is more than doubled that of our models assuming an optimal value for the stellar inclination. In practice, these large \kisr\ values suggest that offsets on the inclination angle in the DI input parameters should remain smaller in real cases. Not surprisingly, statistical error bars are also very large, up to ten times wider than those of our optimal model.

\begin{figure}
\centering
\includegraphics[width=8cm]{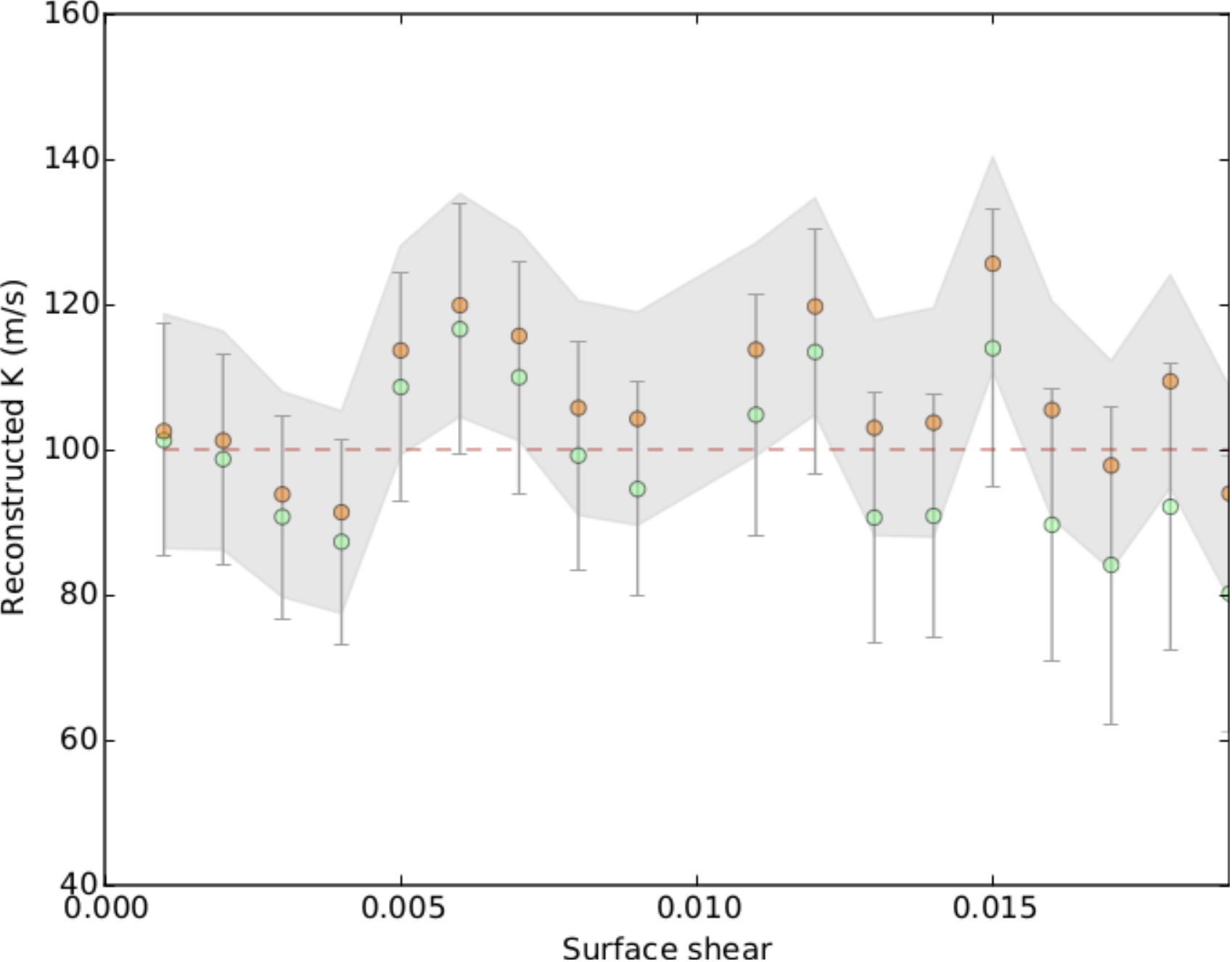}
\caption{Reconstructed $K$ values obtained with stellar models, including various levels of solar-like surface differential rotation. The dimensionless surface shear is equal to $d\Omega / \Omega_{\rm eq}$, where $\Omega_{\rm eq}$ is the rotation rate of the equator and $\Omega$ is the difference of rotation rate between the pole and the equator. Orange dots and their error bars (grey area) correspond to DI models compensating for the differential rotation \citep{petit02}. Green dots are obtained with DI models assuming solid-body rotation.}
\label{fig:drot}
\end{figure}

Finally, we ran a first estimate of the impact of stellar differential rotation on the reconstructed orbital parameters (Fig. \ref{fig:drot}). We assumed here that the stellar surface is altered by a latitudinal shear that follows a simplified law: $\Omega(l) = \Omega_{\rm eq} - \sin^2(l).d\Omega$, where $\Omega(l)$ is the rotation rate at latitude $l$, $\Omega_{\rm eq}$ is the rotation rate of the equator, and $d\Omega$ is the difference of rotation rate between the pole and the equator. We varied the dimensionless parameter $\gamma = d\Omega / \Omega_{\rm eq}$ from zero (solid-body rotation) to 0.02, assuming $(K, P_{\rm orb}/P_{\rm rot}, \phi) = (100~\ms, 1.7, 0.5)$. Assuming a star rotating in three days, $\gamma = 0.02$ roughly corresponds to 80\% of the solar shear level. In this case, our artificial data (assumed to be gathered over three stellar rotation cycles) are collected over nine consecutive days. Some of our DI models compensate for differential rotation (\citealt{petit02}, orange dots). Other DI models simply ignore the shear by assuming solid-body rotation (green dots), showing the consequence of obtaining an incorrect estimate of the shear, or of using an incorrect type of differential rotation law. Fig. \ref{fig:drot} shows that the biases and error bars remain roughly constant for DI models incorporating a differential rotation law. Up to  $\gamma \approx 0.012$, the situation is mostly the same for models assuming solid-body rotation. However,  biases steadily increase past this point whenever the surface shear is ignored. This evolution is coming along with a $\approx 25$\% increase of the \kisr, and a significant increase in the statistical error bars. Above  $\gamma = 0.02$, the majority of our models based on solid-body rotation do not provide any well-defined spottedness minimum in the 3D entropy landscape, so that no reliable values of orbital parameters could be derived.  The two differential rotation parameters can be added to the orbital parameter triplet  to perform a simultaneous optimization of all five parameters. In this case, the additional dimensions to be probed would likely require us to adopt an MCMC approach to identify the best model, instead of the time-consuming systematic paving of a very large 5D grid. Inclusion of the stellar inclination and \vsin\ is possible as well, leading to a 7D space to be explored.

\section{Conclusions}

In spite of a simulated set of line profiles displaying an activity jitter as high as 2.5~\kms\ peak to peak, we are able to robustly recover orbital elements of a fake planet with RV signatures of a few tens of \ms\,, i.e. of the order of magnitude of the RV scatter due to the photon noise of our simulations. This is true as long as the planet is not in a co-rotating configuration, in which case offsets of up to 50\% are observed in the reconstruction of $K$ (assuming $K_{\rm in} = 100$\ms), while \porb\ and $\phi$ are kept closer to their input values.  

Our preliminary tests support the capabilities of this maximum entropy method in a systematic search for hot jupiters around extremely active stars, as long as the collected time series is well adapted for DI, i.e. with a dense rotational sampling collected over a few rotation periods. However, we need to complement the present set of simulations,  with a more comprehensive set of models to determine the detailed effect of a number of parameters (e.g. \vsin, stellar inclination, orbital and rotational phase sampling, surface differential rotation) on the reconstruction of planetary orbits. Varying the inclination will mostly become critical at low $i$ values ($i.e.$ large $K/\sin(i)$ values), since the detectability of low-mass planets will be much reduced in this geometrical configuration. The impact of \vsin\ is more complex to address. The present approach makes use of the fact that, at large projected rotational velocities, the spectral signatures of individual spots are bumps restricted to a fraction of the line profile, while the planetary signature affects the line profile as a whole. Low \vsin\ values will make this disentangling less easy;  the limit of efficiency of the technique is therefore similar to the \vsin\ limit of accurate DI. In any case this threshold is below \vsin\ $= 10$ \kms, since Donati et al., in press, demonstrated that the method is still operative for v819 Tau with \vsin\ $= 9.5$ \kms. On the other hand, if higher \vsin\ values  improve the DI efficiency, at the same time it would  reduce the RV accuracy, so that an upper limit in projected rotational velocity should exist as well.

In addition, the optimization of our detection threshold will likely require the DI reconstruction of both cool spots and faculae \citep{donati14}. Future tests will also assess the benefit of obtaining observations at different epochs to take advantage of the changing spot occupancy to improve the accuracy of planet detection and characterization in repeated measurements. The impact of non-rotationally modulated events that a classical DI code is not able to model, such as flares or limited spot lifetime, will also deserve special attention. 

We also stress that the tests performed here were limited to a single planet following a circular orbit. More complex configurations involving an orbital eccentricity or multiple companions will necessitate probing additional dimensions in a multi-D entropy landscape. Using any real data set, the average RV of the star-planet system is another parameter that cannot be ignored in the multi-D landscape (see Donati et al., in press).

The present maximum entropy method constitutes a very simple way to extract orbital parameters from heavily distorted line profiles of active stars, paving the way towards a systematic search for close-in planets around young, rapidly-rotating stars. It is easily adaptable to most existing DI codes, and may be employed to revisit a number of existing data sets initially collected for DI purpose alone. One specificity of this technique is the simultaneous extraction of a DI map $and$ of orbital elements, in a one-step procedure much simpler than the sequential approach generally proposed in this context. In addition to its simpler approach, the lower amount of data manipulation involved here is likely to make the present strategy less prone to information removal. 

\begin{acknowledgements}
We are grateful to an anonymous referee for the suggestion of a number of new tests, which helped to clarify  the assets and limitations of our method further.
\end{acknowledgements}

%\nocite{*}

\bibliography{jitter}

\end{document}